# Single-photon linear polarimeter based on a superconducting nanowire array

**X. Q. Sun[1,2,3,4], W. J. Zhang[1,3,4†], C. J. Zhang[1,3,4], L. X. You[1,3,4,*], G. Z. Xu[1,3,4], J. Huang[1,3,4], H. Li[1,3,4], Z. Wang[1,2,3,4] and X. M. Xie[1,2,3,4]**

[1]State Key Lab of Functional Materials for Informatics, Shanghai Institute of Microsystem and Information Technology (SIMIT), Chinese Academy of Sciences (CAS), 865 Changning Rd., Shanghai, 200050, P. R. China.

[2]School of Physical Science and Technology, ShanghaiTech University, Shanghai 201210, China.

[3]CAS Center for Excellence in Superconducting Electronics (CENSE), 865 Changning Rd., Shanghai, 200050, P. R. China.

[4]Center of Materials Science and Optoelectronics Engineering, University of Chinese Academy of Sciences, Beijing 100049, China

E-mail: †zhangweijun@mail.sim.ac.cn; * lxyou@mail.sim.ac.cn.

**Abstract:** Superconducting nanowire single-photon detectors (SNSPDs) have attracted remarkable interest for visible and near infrared single-photon detection, owing to their outstanding performance. Conventional SNSPDs are generally used as binary photon-counting detector. Another important characteristic of light, i.e., polarization, has not been resolved using SNSPDs. In this work, we simulated, fabricated, and characterized a linear polarimeter based on a four-pixel NbN superconducting nanowire array, capable of resolving the polarization state of linearly polarized light at the single-photon level. The detector array design is based on a division of focal plane sensor, in which the orientation of each nanowire division (pixel) is offset by 45°. Each single nanowire pixel operates as a combination of photon detector and almost linear polarization filter, with an average polarization extinction ratio of approximately 10. The total system detection efficiency with four pixels is approximately 1% at a total dark count rate of 680cps, when the detector array is free-space coupled and illuminated with 1550nm photons. The Stokes parameters are extracted from polarization measurements of the four pixels. The mean errors of the measured AoP and DoLP were about −3° and 0.12, respectively. Our results indicate that it is possible to develop a scalable polarization polarimeter or imager based on a superconducting nanowire array. This detector array may find promising application in single-photon polarization detection and imaging.

## 1. Introduction

Most of the imaging sensors found in daily life are designed to detect light intensity and wavelength, corresponding to human perception of brightness and color, respectively. Imaging sensors do not generally detect the polarization state of light, since humans are not strongly sensitive to polarization information. However, capturing the optical properties of polarized light has proved to be important, since it can provide additional visual information, even in environments with severe optical scattering, such as in target contrast enhancement in hazy conditions [1], and underwater imaging [2, 3]. Polarization detection and imaging also play key roles in astronomical observations [4, 5], and biomedical imaging studies [6]. Recently, polarization state detection in photon-starved regimes has attracted great interest, as it offers significant improvements in measurement sensitivity, resolution, and accuracy. For example, in 2019, using an advanced photon-counting Lidar system operated at 532nm and 1064nm, NASA's ICESat-2 altimeter achieved unparalleled precision in measuring the heights of Earth's surfaces [7].

Generally, the polarization characteristics of light are described by three important parameters: Light intensity ($I$), angle of polarization (AoP), and degree of polarization (DoP). The most common method for defining these parameters is to use the Stokes vector $S = (S_0, S_1, S_2, S_3)^T$, where superscript $T$ is the matrix transpose, and $S_i$ ($i = 0, 1, 2, 3$) is the Stokes parameter. Stokes vector analysis requires measurement of light intensity at four different polarization angles using a polarimeter. Depending on the time or spatial division configurations used, polarimeters are usually classified as one of three types: The division of time polarimeter (DoTP), division of amplitude polarimeter, and division of focal plane (DoFP) polarimeter [8]. Further details and comparisons of these polarimeters can be found in previous reviews [9, 10].

To analyze the Stokes vector of the incoming light at the photon-starved or even the single-photon level, a common approach is to place a rotating polarizer before the semiconductor single-photon detectors [4], i.e., the DoTP configuration. However, this approach has the disadvantage of requiring the target and detector to be stationary during signal acquisition, making it unsuitable for recording fast moving objects [9, 10]. Moreover, the performance of this type of polarimeter is highly dependent on the performance of the semiconductor single-photon detectors used. For instance, at near infrared wavelengths, a commercial InGaAs/InP avalanche photodiode [11] is generally limited by low system detection efficiency (SDE, ~20%), long dead time (~2μs, free-running), and a high likelihood of afterpulsing.

In recent years, superconducting nanowire single-photon detectors (SNSPDs) have attracted significant attention for their exceptional performance characteristics, including high SDE (>90%) [12, 13], low dark count rate (DCR, <1Hz) [14, 15], low timing jitter (TJ, < 20ps) [16, 17], and high maximum count rate (MCR, > 500MHz) [18]. These detectors have been demonstrated successfully in various applications, including long-distance quantum key distribution [19-21], quantum optics [22], biofluorescence imaging [23], and laser ranging and imaging [24, 25]. SNSPDs are typically single-mode fiber (SMF) coupled, and generally used as binary photon-counting detectors. Their detection mechanism is the generation and disappearance of a photon-induced resistive domain across a superconducting nanowire upon absorptance of a photon [26]. As a result of their meandered nanowire structure, conventional SNSPDs have significantly higher absorption for the transverse electric (TE) mode versus the transverse magnetic (TM) mode. The absorption ratio between TE and TM modes is defined as the polarization extinction ratio (PER), used evaluate the polarization sensitivity of SNSPDs. Much previous research [27-30] has focused on eliminating the polarization sensitivity of SNSPDs. For example, Verma et al [28] fabricated a three-dimensional polarization-insensitive SNSPD by vertically stacking two WSi$_x$ SNSPDs, with a maximum SDE of 87% at 1540nm wavelength and a low PER of 1.04. Conversely, SNSPDs' high polarization sensitivity suggests potential application in polarization detection and imaging. In 2014, Guo et al. fabricated a polarization-sensitive SNSPD that obtained a PER of ~22 and SDE of ~12% with a low filling factor design [31]. In 2018, Xu et al. reported a SNSPD with high PER of ~420 and SDE of ~48%, achieved by placing an Al grating between the Au mirror and NbN meander nanowire [32]. Recently, an emerging multipixel SNSPD array platform [33, 34] has provided new opportunities for developing single-photon sensitive, fast, and large-scale integrated polarization polarimeters or imagers. However, to the best of our knowledge, a single-photon polarimeter based on superconducting nanowires has not previously been reported.

To reach this goal, two major challenges must be addressed: Firstly, it is difficult to resolve the polarization state of light using only one SNSPD, due to the complexity of the Stokes vector. Secondly, the cross-coupling of optical power between polarization modes caused by the birefringence effect [35] in the fiber makes it difficult to calibrate the polarization state of light that actually reaches the detector. Specifically, the polarization state of a SMF-coupled SNSPD varies randomly, because the inherent birefringence varies with external perturbations in the fiber (caused by e.g., bending, twisting, or thermal fluctuations during measurement). In this case, one may consider replacing the SMF with a polarization-maintaining fiber (PMF), which can maintain linear or circular polarization along the fibers. However, use of a PMF prevents the future possibility of measuring any other light polarization states.

To address the problems mentioned above, we simulated, fabricated, and characterized a polarization-sensitive four-pixel NbN SNSPD array, used as a DoFP sensor. The nanowire was patterned with a width of ~50nm and a pitch of ~200nm, to ensure a high average PER of approximately 10. The SNSPD array was characterized with a free-space coupling setup, guaranteeing reliable polarization analysis and real-time calibration in the same optical path. By means of Stokes parameter measurements, we successfully demonstrated a single-photon linear polarimeter based on a superconducting nanowire array for the first time. The absence of moving parts enables the possibility of developing a high performance, real-time polarization detector. Such a detector may be of interest in single-photon polarization detection and imaging applications.

## 2. Concept

### 2.1 DoFP design for SNSPD

In general, the polarization state of light will be transformed after the light linearly interacts with an optical system (the sample). The transformation process is described by

$$S_{out} = MS_{in} \tag{1}$$

where $S_{in}$ and $S_{out}$ are the light's input and output Stokes vectors, and M is the sample's Mueller matrix [36]. The Mueller matrix is a $4 \times 4$ real matrix that characterizes the interaction of the input polarization state with the sample. Using the Mueller matrix, we can obtain all the sample's polarization-changing properties.

To map polarization state using a superconducting nanowire (or SNSPD), we introduce a method analogous to SNSPD. Since the SNSPD is polarization-sensitive, we treat it as a combination of an analog polarizer (AP) and a polarization-insensitive detector (PID), as shown in Fig. 1(a). The transmission axis of the AP (whose angle is denoted by $\theta$, with respect to an x-axis) is parallel to the orientation of the nanowire. Photons transmitted through the AP will be detected by the PID; the most commonly detected will be those polarized in parallel with the nanowire. Based on this hypothesis, we can apply existing theories developed for polarizers to the SNSPD. The theoretical details can be found in Appendix A.

To obtain the optical properties of the input polarized light, according to Stokes' theorem, three important parameters must be measured: $I$, AoP, and DoP. Here the DoP can be divided into degree of linear polarization (DoLP), and degree of circular polarization (DoCP). Since DoFP sensors without a 90° phase retarder do not normally detect circular and elliptical polarization, linearly polarized light is thus incident to the sample in this study, and the Stokes parameter $S_3 = 0$ (i.e., DoCP = 0). In low light levels, the intensity $I$ is proportional to the number of photons per unit of time received by a single-photon detector, i.e., the detector's photon count rate (PCR).

To obtain the Stokes parameter $S_i$, we must measure $I_\theta$ or $PCR_\theta$. Then, we construct the DoFP layout using the SNSPD, as shown in Fig. 1(b). A superpixel is composed of four pixels with nanowires oriented at $\theta$ = 0°, 45°, 90°, and 135°, respectively. The corresponding PCR of the nanowire is denoted by $PCR_\theta$. If there are suitable cryogenic readout circuits, the layout of the DoFP structure can be easily expanded to a large-scale array, e.g., a 16-pixel array, as shown in Fig. 1(c).

We can express $S_i$ in terms of $PCR_\theta$:

$$S_0 = PCR_{0°} + PCR_{90°} \qquad (2\text{-}1)$$

$$S_1 = PCR_{0°} - PCR_{90°} \qquad (2\text{-}2)$$

$$S_2 = PCR_{45°} - PCR_{135°} \qquad (2\text{-}3)$$

Notably, in our study, the incident light is backside illuminated. From the perspective of the incident light, the nanowire orientations at 45° and 135° shown in Fig. 1 (b) will be exchanged, while the nanowire orientation of 0° and 90° will be unchanged. Using the above relationships, we have

$$S_2 = PCR_{135°} - PCR_{45°} \qquad (2\text{-}4)$$

Therefore, according to the definitions of AoP and DoLP [9],

$$AoP = \frac{1}{2}\arctan\left(\frac{S_2}{S_1}\right) \qquad (3\text{-}1)$$

$$DoLP = \frac{\sqrt{S_1^2 + S_2^2}}{S_0} = \sqrt{\left(\frac{S_1}{S_0}\right)^2 + \left(\frac{S_2}{S_0}\right)^2} \qquad (3\text{-}2)$$

combined with Eq. (2-1), (2-2), and (2-4), we can represent AoP and DoLP in terms of $PCR_\theta$,

$$AoP = \alpha = \frac{1}{2}\arctan\left(\frac{PCR_{135} - PCR_{45°}}{PCR_{0°} - PCR_{90°}}\right) \qquad (4\text{-}1)$$

$$DoLP = \frac{\sqrt{(PCR_{0°} - PCR_{90°})^2 + (PCR_{135°} - PCR_{45°})^2}}{PCR_{0°} + PCR_{90°}} \qquad (4\text{-}2)$$

Here, we denote AoP as $\alpha$. The Stokes parameters are often normalized to the value of $S_0$, such that the differences between $I_\theta$ and $PCR_\theta$ will be eliminated in the process of solving the equations for the DoLP and AoP.

It is known that the PCR of the SNSPD is proportional to its SDE ($\eta_S$) when the incident photon rate ($R_{in}$) is constant (or calibrated), i.e.,

$$PCR = \eta_S R_{in} \qquad (5)$$

Generally, $\eta_S$ can be expressed [12, 13] as

$$\eta_S = \eta_C \cdot \eta_A \cdot \eta_I \qquad (6)$$

where $\eta_C$ is the optical coupling efficiency, $\eta_A$ is the absorption efficiency, and $\eta_I$ is the intrinsic detection efficiency of the nanowire. According to Eq. (5) and (6), we have

$$PCR = \eta_C \cdot \eta_A \cdot \eta_I \cdot R_{in} \qquad (7)$$

For each run of the SDE measurement, the $\eta_C$ is constant, and the SNSPD will obtain a maximum absorption efficiency ($\eta_{A\parallel}$) or maximum PCR ($PCR_{max}$) when the light polarization is parallel to the nanowire (i.e., TE mode), and a minimum absorption efficiency ($\eta_{A\perp}$) or minimum PCR ($PCR_{min}$) when the light polarization is perpendicular to the nanowire (i.e., TM mode). Then, we can further define PER in terms of PCR or $\eta_A$ as

$$PER = \frac{PCR_{max}}{PCR_{min}} = \frac{\eta_{A\parallel}}{\eta_{A\perp}} \qquad (8)$$

## 2.2 SNSPD Finite element simulation

We simulated the optical absorption of the nanowire on a SiO$_2$/Si/SiO$_2$ substrate (268nm SiO$_2$, 400μm Si) by varying the nanowire's geometrical parameters. The simulation was performed using commercial finite element software (Comsol Multiphysics, RF module) [31]. Figure 1 (d) shows a cross-sectional schematic of the nanowire used in the simulation. To minimize computation time, the thickness of the Si layer in the simulation was set to 2μm. Figure 1(e) shows the simulated optical absorptance for TE and TM modes, and PER as a function of the width, with nanowire pitch and thickness fixed at 200nm and 7nm, respectively. As expected, $\eta_{A\parallel}$ (TE mode, red circular dots) was higher than $\eta_{A\perp}$ (TM mode, blue square dots) for the same nanowire width. The PER continuously increased with reduction in nanowire width. For 100nm and 50nm wide nanowires, the PER was 3.6 and 16.8, respectively.

Since a high PER is preferable in the present study, a detailed investigation of the high PER nanowire structure was necessary. For this, we selected a 50nm wide, 7nm thick, 200nm pitch nanowire as an example, since was feasible to fabricate with a standard process. Figure 2(a) shows $\eta_{A\parallel}$, $\eta_{A\perp}$, and PER for the 50nm wide nanowire as a function of the incident angle of light. When the light was normally incident to the nanowire (i.e., 0°), $\eta_{A//}$ and $\eta_{A\perp}$ were approximately 25% and 1.5%, respectively, meaning that the PER was approximately 16.8. In the incident angle range of 0°–20°, $\eta_{A}$ and PER were almost unchanged. However, as the incident angle increased, $\eta_{A//}$ gradually increased and reached a maximum of ~65% around 80°, where a total internal reflection appeared at the interface between the top $SiO_2$ layer and the air layer. As the incident angle further increased to 90°, $\eta_{A//}$ dropped quickly. Therefore, the increase in $\eta_{A//}$ with the increase in incident angle (<80°) was due to the cavity-enhanced absorption of the reflected light from the $SiO_2$/air interface, since the nanowire was embedded in a quarter-wavelength cavity and the intensity of the reflected (transmitted) light increased (decreased) with the increase in incident angle. Moreover, $\eta_{A\perp}$ fluctuated slightly. Under the joint action of $\eta_{A//}$ and $\eta_{A\perp}$, PER continuously increased to a maximum of ~ 86, (on a log scale) around 57° and then slowly fluctuated with further increase in the incident angle.

Although PER is dependent on incident angle, it is preferable to use normally incident light to simplify the model. In our experiment, a lens was used to focus the incident light at an incident angle estimated to be less than 3°. Thus, the influence of the incident angle on PER was negligible in this study. In Fig. 2(b)–(c), distributions of electromagnetic power loss density (i.e., optical absorption) between TE and TM modes show significant contrast, resulting in a high PER value. Figure 2(d) shows a simulation of absorptance as a function of $\alpha$. The results can be well fitted with a cosine function (a variant of Malus's law of polarization):

$$\eta_A = a \cdot \cos\frac{\pi}{180}\frac{2\alpha}{b} + c \tag{9}$$

where $a = 0.12$, $b = 1$, and $c = 0.13$ are the fitting parameters. The coefficient $\frac{\pi}{180}$ represents conversion between degrees and radians.

## 3. Fabrication and experimental setup

A 7nm thick NbN thin film was deposited by magnetron sputtering onto a two-inch thermal-oxidized Si substrate, with 268nm thick $SiO_2$ on both sides of the substrate. The film was then patterned into a nanowire array using electron beam lithography (EBL, JOEL, 100kV accelerating voltage), and reactive ion etched in $CF_4$ plasma. Each pixel of the array comprised a meandering nanowire structure, with nanowire orientations of 0°, 45°, 90°, and 135°, respectively. Fig. 3 presents a typical scanning electron microscopy (SEM) image of the array, with a nanowire width of 45 ± 5nm and pitch of 200 ± 5nm for each pixel. The SEM images indicate that good nanowire uniformity was obtained in fabrication. The total active area of the array was $16 \times 16 \mu m^2$.

We first screened the fabricated devices in a 2K Gifford–McMahon (G–M) cryocooler system, based on the uniformity of the switching currents ($I_{SW}$) and the saturation of the current-dependent SDE for all pixels in the array. We defined $I_{SW}$ as the highest bias current that the device could sustain before switching to the normal state. The selected device was then mounted in a 1.5K free-space-coupled cryogenic system (Cryomech Inc., PT410) for further characterization, since testing in a free-space optical path can avoid the drawbacks of SMF. Figure 4 shows a schematic diagram of the experimental setup. The inset of Fig. 4 presents a photo of the chip package for an SNSPD array chip, which can support up to 16 channels of signal output.

As shown in Fig. 4, light emitted from a 1550nm femtosecond laser (Calmar laser, FPL-01CF) was transmitted through a SMF and then fed into two cascaded variable fiber attenuators (Keysight, 81576A). After undergoing strong attenuation to the single-photon level, the light was collimated and converted into free space using a fiber collimator (Thorlabs, PAF2S-18C). In free space, the light was linearly polarized to a specific state using a linear polarizer (Thorlabs, LPNIR100-MP2, extinction ratio up to $10^5$). The polarized light passed through a zero-order half-wave plate (Thorlabs, WPH10ME-1550, ±1° step accuracy), which could be rotated to control the state of polarized light. A movable commercial semiconductor polarimeter (Thorlabs, PAX1000IR2/M, dynamic range of −60 to +10dBm, accuracy of ±0.25°) was used to calibrate the polarization state. A movable optical power meter (Thorlabs, S132C, ±5% measurement uncertainty, 9.7 × 9.7 mm$^2$ active area) was used to calibrate the input power of the incident light. Before each polarization measurement, the laser power was calibrated at a high light level to avoid PCR changes due to variations in input power. The calibrated light propagated to the device via two optical filters (placed at the 40K and 4K shields of the cryostat) and a lens. The lens was attached to the package block of the device (cooled at the 1K stage), to focus the light spot to a size of about 20μm in diameter. With the lens in place, the SDE was greatly improved and the DCR was also significantly reduced, since the lens components were designed to transmit 1550nm light with a bandwidth of ± 5nm, while filtering stray light in other bands. The insertion loss for the two filters measured at room temperature was about 0.4dB. The loss due to beam divergence (not fully collected by the lens) was approximately 0.1dB. Thus, the total transmission loss from the fiber output port (before connecting to the fiber collimator) to the device was approximately 0.5dB.

The nanowire array was illuminated from the back side of the substrate by light transmitted through free space. The half-wave plate controlling the polarization angle was rotated to obtain maximum (minimum) count rates for each pixel, corresponding to the polarization direction parallel (perpendicular) to the nanowire. At these specific angles, the SDE of each pixel was recorded as a function of bias current. By blocking the laser with a shutter, we recorded the count rates as the DCR for each pixel. The total SDE and DCR of the multipixel array was obtained by summing the SDE and DCR values recorded for each pixel. Because the light was normally incident to the elements of the optical path during the polarization measurements, changes in polarization state were negligible.

## 4. Results and analysis

### 4.1  SNSPD performance in the free-space system

Figure 5(a) shows photon-response pulse waveforms for the four pixels, recorded using an oscilloscope. The pulses are nearly identical, with a pulse-decay time of about 15ns (1/e criterion). The inset of Fig. 5(a) shows $I_{SW}$ distribution for the four pixels. The $I_{SW}$ values are within the range $6.7 \pm 0.4\mu A$ at 1.5K, confirming good uniformity in the fabricated nanowire array. Figure 5(b) shows the typical timing jitter of a single pixel, with a full width at half maximum of 126ps for the histogram, at a bias current ($I_B$) of 5.2µA.

Figure 5(c) shows the SDE of each pixel as a function of $I_B$, when the photons were polarized parallel to the nanowire at a wavelength of 1550nm. All pixels demonstrated a well-saturated SDE plateau (as $I_B \geq 5.0\mu A$), indicating $\eta_i \approx 1$. The maximum SDEs of the four pixels are distributed across the range 0.1%–0.5%, implying a misalignment between the light spot and the four-pixel array. The misalignment could be caused by shrinkage of the lens holder in low temperatures. This shrinkage also could lead to the lens losing focus, creating a larger light spot than expected.

When all the pixels were biased at 5.5µA (in the saturated SDE region), the total SDE of the array was about 1% and the total DCR about 680cps. In other words, the DCR of each pixel was around 170cps at this $I_B$. In our experiment, the measured $PCR_{max}$ for each pixel was of the magnitude $\sim 1 \times 10^4$cps, which is nearly two orders of magnitude higher than the DCR, guaranteeing high contrast during data acquisition.

To explore the upper boundary of this device's SDE, we re-measured it in a 2K G-M cryocooler system by coupling the device with a GRIN lens fiber with backside illumination [31]. The average PER measured using fiber coupling was about 12, which is slightly higher than that measured with free-space coupling (~10). However, the total SDE measured with fiber coupling was about 11%, which is about one order of magnitude higher than the free-space coupled measurement. This result confirms that the relatively low SDE in the free-space system is due to low optical coupling efficiency, since only the optical coupling differed between these two measurements.

**4.2 PCR versus angle of polarization**

Figure 6(a) shows the AoP dependence of the $PCR_\theta$s for all four pixels, measured at $I_B = 5.5\mu A$. $PCR_\theta$s for the four pixels were recorded when the AoP of the incident linearly polarized light was rotated clockwise from $-90°$ to $90°$ in steps of $10°$.

Due to the misalignment between the light spot and the pixel array, the maximum $PCR_\theta$ ($PCR_{\theta max}$) in the $PCR(\alpha)$ function is significant different. Thus, we performed a calibration on these data, by assuming uniform illumination and uniform $\eta_{A//}$ for each pixel. To achieve this, the $PCR_{\theta max}$ of each pixel needs to be identified. This requirement can be realized by normalizing the $PCR_\theta$ of each pixel to its own $PCR_{\theta max}$ in the $PCR_\theta(\alpha)$ function. Combined with Eq. (7), we define a normalized $PCR_\theta$ for each pixel, denoted as $PCR_\theta^*$,

$$PCR_\theta^* = \frac{PCR}{PCR_{max}} = \frac{\eta_C \cdot \eta_A \cdot \eta_I \cdot R_{in}}{\eta_C \cdot \eta_{A\perp} \cdot \eta_I \cdot R_{in}} = \frac{\eta_A}{\eta_{A//}} \tag{10}$$

In this manner, the difference between $PCR_{\theta max}$ of the four pixels is eliminated, while the influence in $\eta_A$ due to variations in $\alpha$ (i.e., the polarization information of the incident light) is preserved.

Figure 6(b) shows the $PCR_\theta^*$ for each pixel as a function of AoP. Due to the limited PER for each pixel, the minimum $PCR_\theta^*$ is about 0.1, which is higher than the ideal value of ~0. The most obvious feature of the figure is that the curves of the four pixels are offset by about ~45° due to the physical orientations of the four nanowires. For quantitative analysis, we fitted the experimental data with cosine functions (indicated with dashed lines). A good fit was obtained with the function expressed below:

$$PCR_{\theta fit}^* = a_0 \cdot \cos\frac{\pi}{180}\frac{2(\alpha - \theta_f)}{b_0} + c_0 \tag{11}$$

where the fitting parameters $a_0 = 0.44 \pm 0.02$, $b_0 = 0.994 \pm 0.01$, and $c_0 = 0.55 \pm 0.01$, and the fitting $\theta_f = 6.2°$, $45.9°$, $94.9°$, and, $139.4°$ differ slightly from the design values (0°, 45°, 90°, and 135°). This may be due to misaligned orientation between the transmission axis of the half-wave plate and the pixel nanowire. This angular-shift error is systematic and could be removed with further calibration. The fitting functions for PCRs are similar to the function for theoretically predicting $\eta_A$, shown in Eq. (9), indicating that the simulation works well for the present case.

### 4.3 Analysis of the polarization angle

We calculated the AoP by substituting the $PCR_\theta^*$ shown in Fig. 6(b) into Eq. (4-1). The measured AoP data are shown in Fig. 7(a), plotted as a function of the reference AoP. The reference AoP was recorded using a commercial polarimeter at a high light level. We performed the AoP measurement twice, obtaining two sets of results, denoted data-A and data-B. Data-A(B) is well fitted with a line of slope 0.98 ± 0.01 (0.97 ± 0.01), a Y-intercept of 4.1 ± 0.3 (1.9 ± 0.6), and an R-square coefficient of 0.999 (0.997). The slight difference in the Y-intercept of the fit line between the two measurements is caused by limited half-wave plate rotation accuracy. By calculating the average differences between the measured AoP and the reference values, the mean AoP error obtained for the two measurements is about −3°. The minus sign here indicates that the measured value is smaller than the reference value. We corrected the AoP error by adding an angular offset of about 4.1° (i.e., the fitted Y-intercept) to the data-A measurements. Figure 7(b) shows the corrected AoP versus the reference AoP. Clearly, the corrected curve almost overlaps an ideal function y = x (orange line in the figure). This result illustrates that the angular-shift error contributes to the AoP error, and can be reduced in the manner described.

### 4.4 Impact of PER on the measurement of Stokes parameters

To study the effect of PER on the accuracy of the polarization state measurement, we extracted the Stokes parameters of the incident light using a different device with a lower average PER value of about 3 [see Appendix C], comprised of a 100nm wide nanowire array with the same pitch and thickness as the 50nm wide device used in the present study.

We then performed an experiment to compare these two devices. Figure 8 shows the measured Stokes parameters ($S_1$, and $S_2$), AoP, and DoLP as a function of the sampling sequence number. Reference data obtained from the commercial polarimeter (red dots) are also plotted in the figure. As expected, the 50nm wide (high PER) SNSPD demonstrates better agreement with the reference values than the 100nm wide (low PER) device. In detail, the measured $S_1$, $S_2$, and DoLP values demonstrate significant sensitivity to PER value. We will use the measured DoLP value as an example, since it is most sensitive to the changes in PER values, due to the DoLP (Eq. (4-2)) calculations involved in all measured $PCR_\theta$. Theoretically, DoLP = 1 indicates an ideal linearly polarized state. As shown in Fig. 8(a) and (b), the reference DoLP value is ~1, indicating that on-demand generation and control of the linearly polarized state was achieved in our experiment. For a device with PER of ~10 or 3, the measured DoLP has a mean value of 0.88 or 0.65, respectively. Correspondingly, the mean error of the measured DoLP is about 0.12 or 0.35, respectively. Thus, a higher PER value would guarantee a more accurate measured DoLP. Meanwhile, the measured and reference AoP data almost overlap in both cases, i.e., the measured AoP values are less sensitive to change in PER values, owing to the significant difference between the maximums and minimums in the $PCR(\alpha)$ function. The 100nm wide nanowire array has mean AoP error of about −5.5°, which is slightly larger than the 50nm wide device. However, even for the case with PER of ~10, errors appear in the measured $S_1$, and $S_2$, due to the limited PER and the simplified model used. These errors could be eliminated in the future by use of a higher PER value or additional calibration.

## 5. Discussion

In this study, an assumption of uniform illumination and uniform $\eta_{A/\!/}$ was made to calibrate the measured $PCR_\theta$, owing to inefficient optical coupling. This assumption ignores certain information, such as differences in active area, wire width, and coupling efficiency between the nanowire pixels. $PCR_\theta$ normalized under this assumption produces acceptable experimental results, which indicates that either the differences mentioned above are quite small, or these physical quantities contribute less error than the others (such as the PER). We speculate that the PER could be a dominant factor in the simplified model. Moreover, this assumption can be relaxed by improving the coupling accuracy or using uniform illumination.

An SNSPD-based polarimeter offers several advantages: Firstly, due to the combination of polarization sensitivity and single-photon detection ability, a SNSPD-based polarimeter can avoid misalignment issues introduced by the integrated charge-coupled device (CCD) polarization image sensor fabrication process [8], in which the micro polarizer array must be accurately stacked on the top of the CCD array. Secondly, the multifunctionality of an SNSPD-based polarimeter would greatly improve system integration, particularly in the fields of quantum communication and remoting sensing, where polarization coding is required. Thirdly, due to the inherent advantages of the SNSPD itself, SNSPD-based polarimeters promise a high SDE and low DCR, as well as low dead time and low timing jitter.

Future development of a high performance, scalable SNSPD-based polarimeter presents multiple opportunities and challenges. Firstly, the SDE of the detector in this study is limited by low optical coupling and absorption efficiency. However, encouraging results have shown that by integrating Al gratings, a single pixel SNSPD can achieve simulated [37] and experimental [32] efficiencies of ~95% and ~48%, respectively, for parallel polarization with a PER of ~1.5 × $10^4$, and ~420, respectively. Thus, the development of a high SDE, high PER, SNSPD-based polarimeter is feasible with proper optical design. Conversely, previous studies of semiconductor DoFP sensors have shown that the high PER requirement can be relaxed if careful calibration is applied to the Mueller matrix, e.g., Eq. (12) in Appendix A. Specifically, it has been shown that a PER of ~3 is sufficient for polarimetry, although PER >10 is preferable for accurate polarization reconstruction [38]. Thus, a DoFP sensor comprising a conventional SNSPD pixel (typical PER ~3–4 [13]) may also correctly resolve the polarization state, making the fabrication of a high performance SNSPD-based polarimeter more feasible. Secondly, increasing the scale of the array would not only help to improve coupling efficiency, but may also enable realization of a polarimetric imager. Of course, a large-scale array would require more complex calibration, e.g., elimination of sampling error due to the instantaneous fields of view of neighboring pixels [39]. However, some of these issues have already been addressed in studies of semiconductor DoFP sensors. Thirdly, integrating a phase retarder could enable the detector to resolve circular polarization, facilitating development of an SNSPD-based full Stokes parameter polarimeter.

The scale expansion of the SNSPD-based polarimeter is limited by the number of cryogenic coaxial cables installed, which would increase the heat load on the refrigerator. This problem is also one of the challenges encountered in the development of multipixel SNSPD arrays. It is encouraging that recently the SNSPD array and cryogenic readout technology have developed rapidly [33, 34, 40-42]. For example, NIST successfully demonstrated a 32 × 32-pixel SNSPD imaging array [33] based on row-column readout architecture, which reduces the number of the readout cable from $32^2$ to 64. NICT also recently reported a 64-pixel array [34] based on a single-flux-quantum circuit, which reduces the number of readout cables by adopting cryogenic digital multiplexing. Obviously, these techniques can be directly applied to our SNSPD-based polarimeter. To some extent, the scale of 1024 pixels is sufficient to support the requirements of small-scale imaging applications. Furthermore, the reported SNSPD imaging array [33] has demonstrated a maximal counting rate of ~900Mcps and a timing jitter of 250–400ps. These performances are better than those of current CCD image sensors [5]. With the development of readout technology in the future, the scale expansion of the SNSPD-based polarimeter is promising.

## 6. Conclusion

In summary, we present a linear polarimeter based on an array of SNSPDs, including simulation, fabrication, and experimental verification. The pixel of the array comprises a ~50nm-wide, 0.25 filling factor NbN nanowire, with an average PER of ~10. The four nanowires were oriented with an offset of 45°, forming a division of the focal plane. By characterizing the device in a free-space coupling cryostat, we confirmed successful detection of the polarization state of a linearly polarized light at 1550nm, with a total SDE of ~1% at a low total DCR of ~680cps. The mean errors of the measured AoP and DoLP were about −3° and 0.12, respectively. Our result indicates that it is feasible to develop a high performance, scalable, single-photon sensitive polarimeter based on SNSPDs. Utilizing the remarkable advantages of SNSPD devices, our SNSPD-based single-photon linear polarimeter may have promise in multiple applications, such as astronomical observations, biomedical imaging, and polarimetric Lidar.

**Acknowledgments**

This work is supported by the National Natural Science Foundation of China (NSFC, Grant No. 61971409), the National Key R&D Program of China (Grant No. 2017YFA0304000) and the Science and Technology Commission of Shanghai Municipality (Grant No. 18511110202). Shanghai Municipal Science and Technology Major Project (Grant No.2019SHZDZX01). W. J. Zhang is supported by the Youth Innovation Promotion Association (No. 2019238), Chinese Academy of Sciences.

**Figure caption**

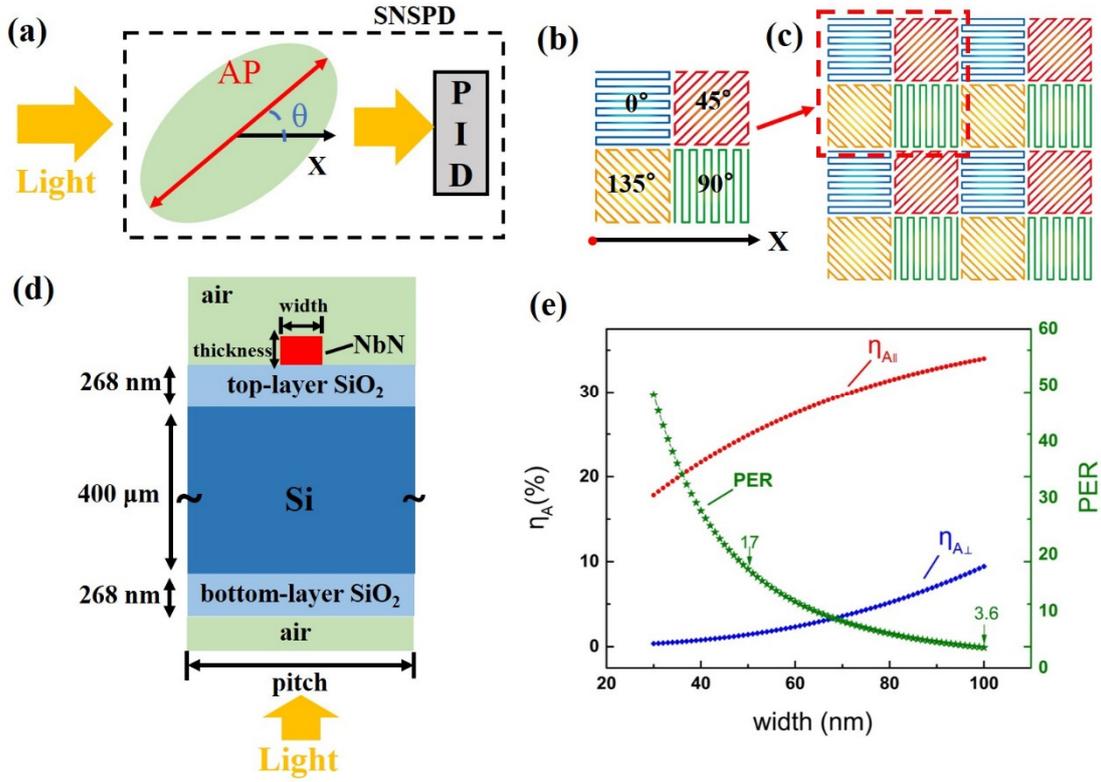

Figure 1. Illustration of the operating principle of a superconducting nanowire array-based polarimeter. (a) Schematic of the SNSPD interaction with incoming light. The SNSPD can be viewed as a combination of an analog polarizer (AP) and a polarization-insensitive detector (PID). A polar coordinate system is used, where $\theta$ is the transmission axis of the AP with respect to an x-axis. The orientation of the nanowire is parallel to the AP's transmission axis; (b) A superpixel of the array comprises four nanowire pixels oriented at $0°$, $45°$, $90°$, and $135°$, respectively; (c) Schematic of a 16-pixel DoFP structure. The four different colors represent the four different nanowire orientations; (d) Cross-sectional schematic of the nanowire on a thermally oxidized Si substrate. Incident light is backside illuminated through the substrate to the nanowires; (e) Simulated optical absorptance $\eta_A$ ($\eta_{A\parallel}$ for TE mode, $\eta_{A\perp}$ for TM mode) and PER as a function of nanowire width, for a simulated nanowire of 7nm fixed thickness and 200nm fixed pitch.

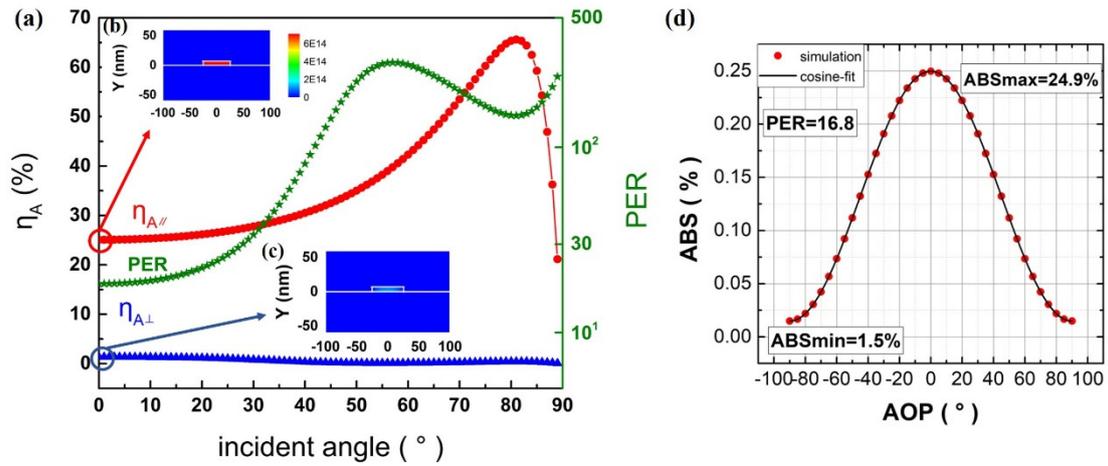

Figure 2. (a) Simulated optical absorptance $\eta_A$ ($\eta_{A//}$ and $\eta_{A\perp}$) and PER as a function of the incident angle of light. The PER is shown on a logarithmic scale. Electromagnetic power loss density distribution for two different modes of normal incident light: (b) TE mode; (c) TM mode; (d) Simulation of the absorptance as a function of the angle of polarization. The red line is a cosine function fit. A 50nm-wide, 200nm pitch, 7nm-thick nanowire was used in the simulation.

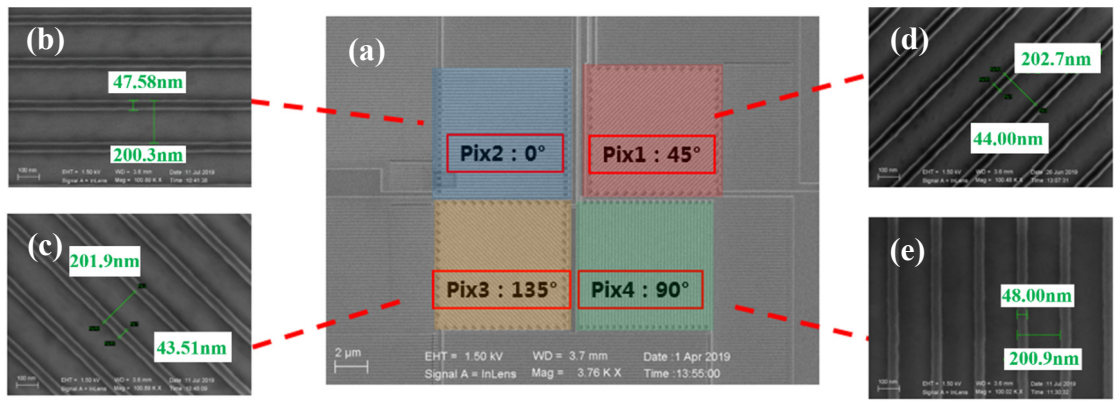

Figure 3. (a) SEM image of a superpixel of the division of focal plane (DoFP). Four different nanowires are oriented with a 45° offset. Magnified SEM images of each pixel: (b) Pix1; (c) Pix2; (d) Pix3; and (e) Pix4. All four pixels show a uniform width of 45 ± 5nm and a pitch of ~200nm.

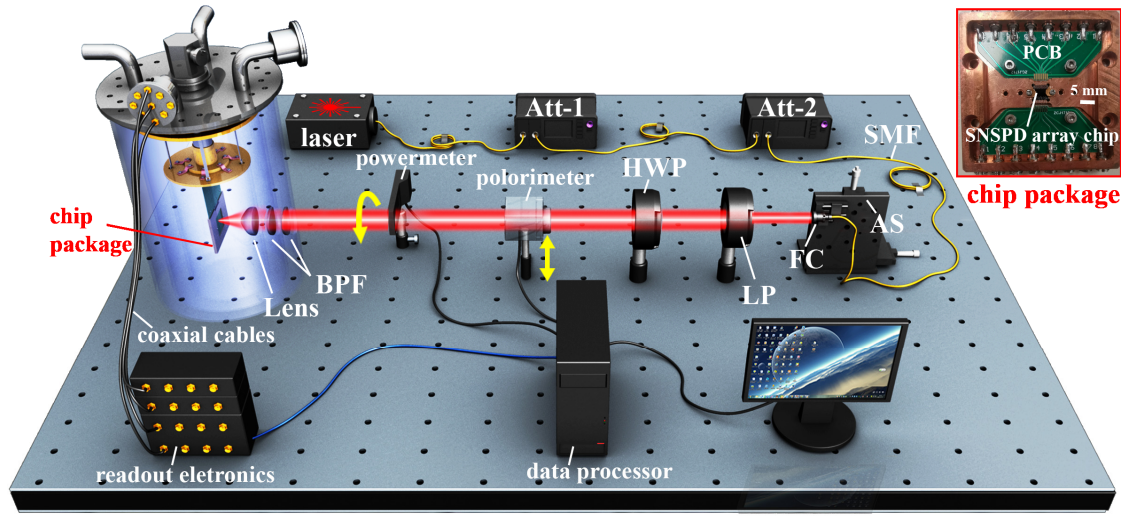

Figure 4. Schematic of the experimental setup of a free-space coupling SNSPD array used for polarization analysis and calibration, and for SDE measurement. Att-1/2: variable optical attenuator; SMF: single-mode fiber; AS: alignment stage; FC: fiber collimator; LP: linear polarizer; HWP: half-wave plate; BPF: bandpass filter. Inset: optical photo of the SNSPD array chip package (front side). The chip size is $5 \times 5$ mm$^2$. PCB: printed circuit board.

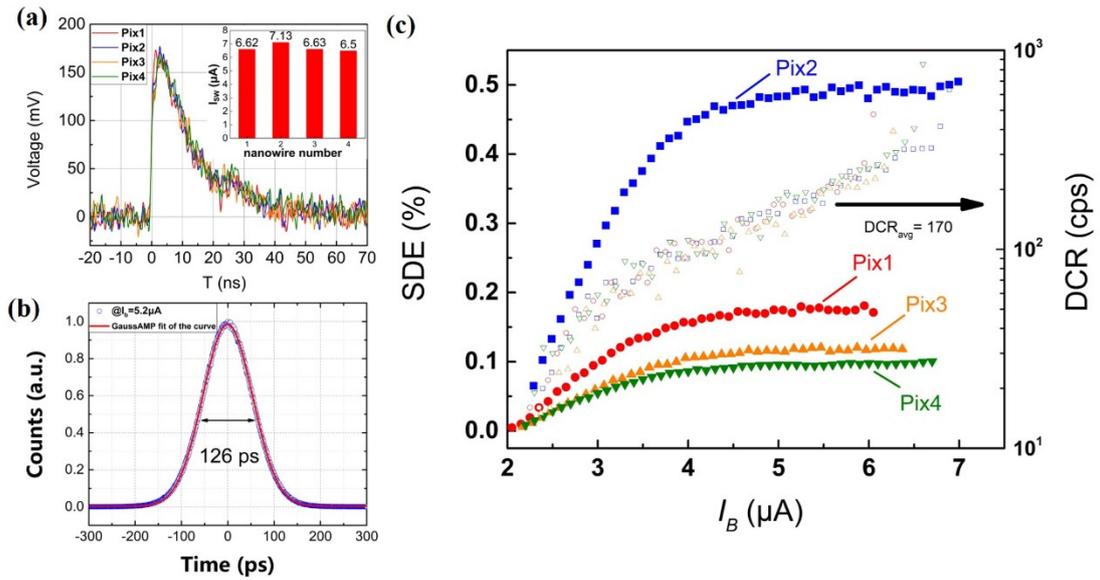

Figure 5. (a) Single-shot waveforms of the photon-response pulse for the four SNSPD pixels, recorded at a bias current ($I_B$) of 5.5μA. Inset of (a), switching current ($I_{SW}$) distribution of the four-pixel array; (b) Timing jitter of a single pixel with a 50nm wide nanowire, measured at $I_B$ = 5.2μA. The red line is the Gaussian fit for the experimental data; (c) SDE (solid scatters) and DCR (open scatters) of each pixel as a function of $I_B$.

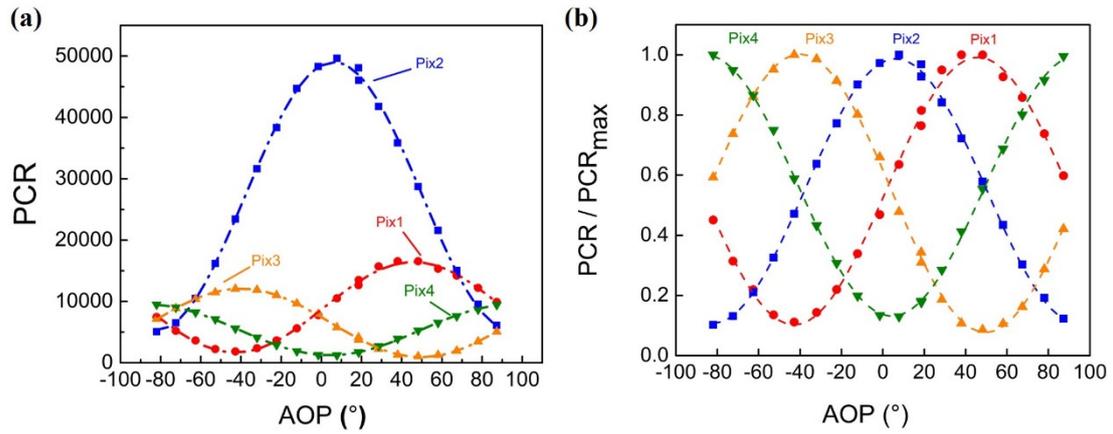

Figure 6. PCR of each pixel as a function of the angle of polarization (AoP). (a) raw data of the PCR$_\theta$; (b) normalized PCR$_\theta^*$ (PCR$_\theta^*$=$\frac{PCR_\theta}{PCR_{\theta max}}$). The dashed lines are cosine function fits. The PCR$_\theta^*$ maximum and minimum values for the four pixels exhibit a ~45° shift from each other along the AoP axis.

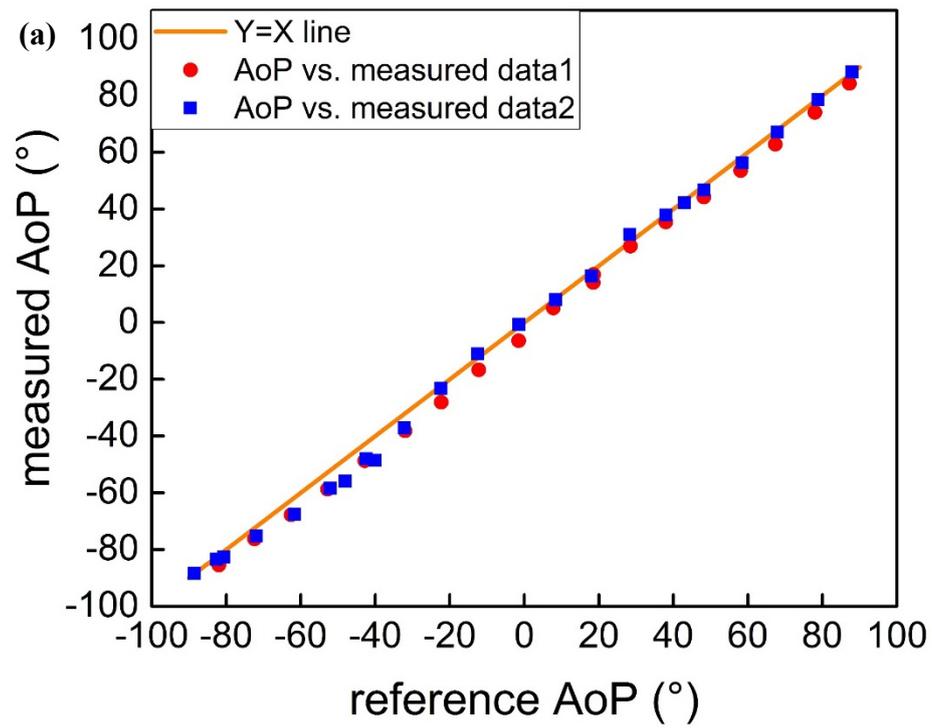

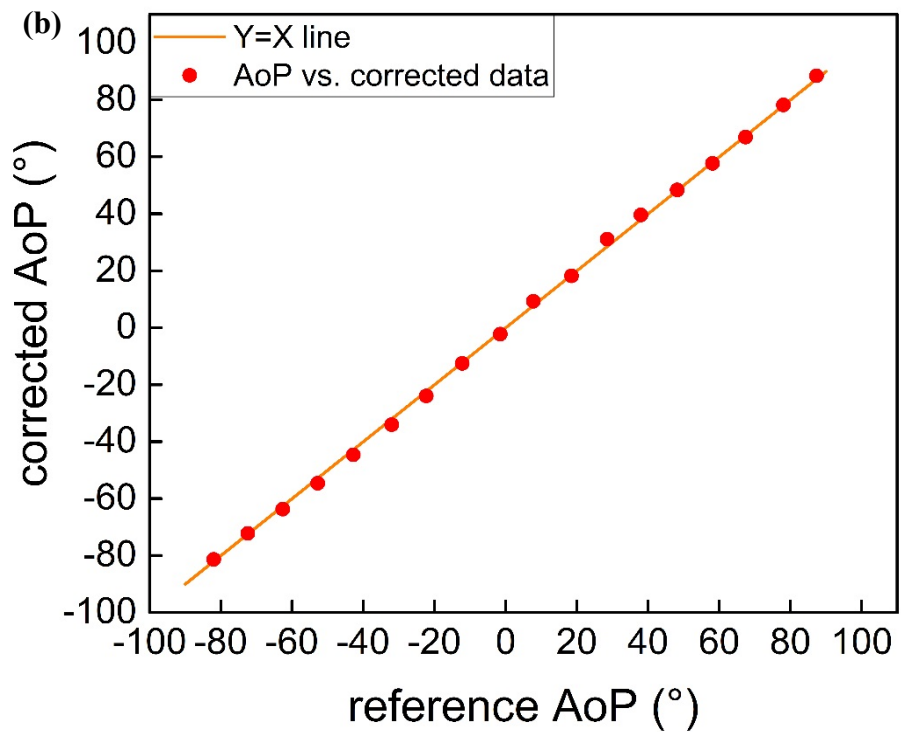

Figure 7. (a) Measured AoP vs. reference AoP. The orange line is an ideal function (x = y) used for indication purposes; (b) Corrected AoP vs. reference AoP.

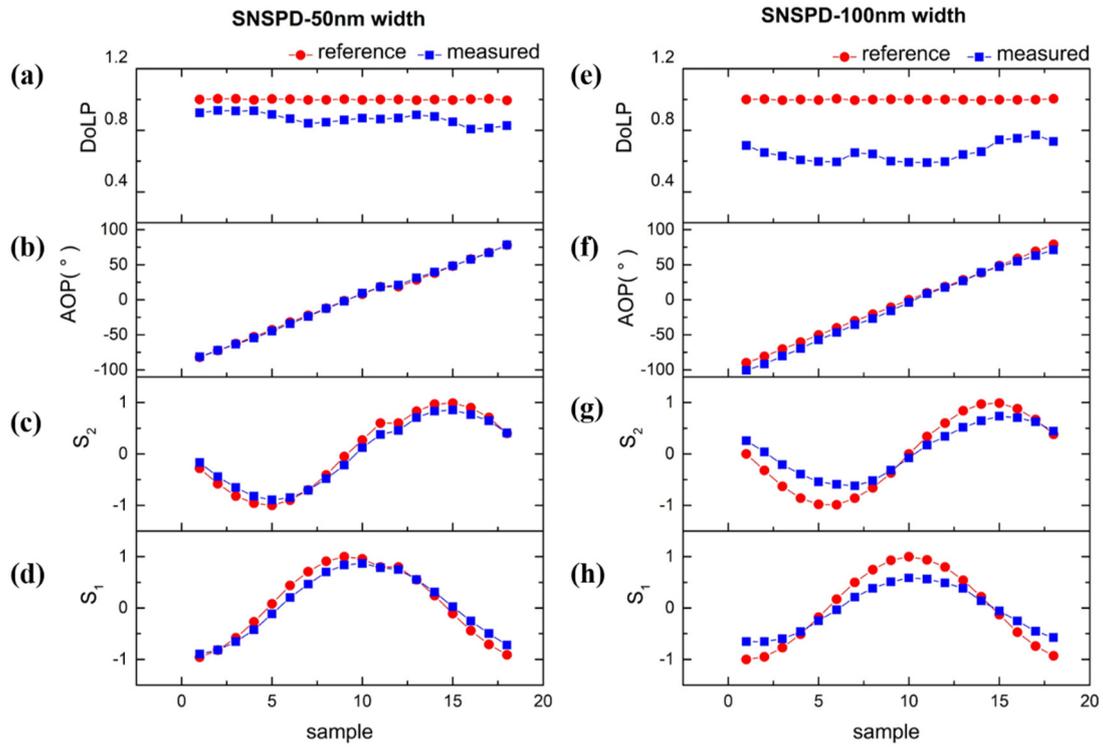

Figure 8. Measured DoLP, AoP, and Stokes parameters $S_1$ and $S_2$ as a function of sampling sequence number, extracted using two devices with different PER values: (a)–(d) 50nm wide nanowire array with an average PER of about 10; (e)-(h) 100nm wide nanowire array with an average PER of about 3. The reference data in each case are plotted with red dots.

## Appendix A: Theoretical model deduced from a polarizer's Mueller matrix

A partial polarizer's Mueller matrix $M_P$ [36] is represented by

$$M_P = \frac{1}{2} \begin{bmatrix} q+r & (q-r)cos2\theta & (q-r)sin2\theta & 0 \\ (q-r)cos2\theta & (q+r)cos^2 2\theta + 2\sqrt{qr}sin^2 2\theta & \frac{1}{2}(\sqrt{q}-\sqrt{r})^2 sin4\theta & 0 \\ (q-r)sin2\theta & \frac{1}{2}(\sqrt{q}-\sqrt{r})^2 sin4\theta & 2\sqrt{qr}cos^2 2\psi + (q+r)sin^2 2\theta & 0 \\ 0 & 0 & 0 & 2\sqrt{qr} \end{bmatrix} \quad (12)$$

where $q$ is the major transmittance, and $r$ is the minor transmittance. As $M_P$ shows, calculations for a non-ideal polarizer would be complex. For an ideal linear polarizer (iLP), i.e., $q = 1$, $r = 0$, $M_P$ can be reduced to $M_{iLP}$.

$$M_{iLP} = \frac{1}{2} \begin{bmatrix} 1 & cos2\theta & sin2\theta & 0 \\ cos2\theta & cos^2 2\theta & \frac{1}{2}sin4\theta & 0 \\ sin2\theta & \frac{1}{2}sin4\theta & sin^2 2\theta & 0 \\ 0 & 0 & 0 & 0 \end{bmatrix} \quad (13)$$

$M_{iLP}$ corresponds to the simplest case for $M_{AP}$, i.e., when the PER of the SNSPD is much greater than 1. Then, we obtain four Mueller matrices corresponding to each nanowire orientation $\theta = 0°$, $45°$, $90°$, and $135°$.

$$M_{AP0} = \frac{1}{2} \begin{bmatrix} 1 & 1 & 0 & 0 \\ 1 & 1 & 0 & 0 \\ 0 & 0 & 0 & 0 \\ 0 & 0 & 0 & 0 \end{bmatrix} \qquad M_{AP45} = \frac{1}{2} \begin{bmatrix} 1 & 0 & 1 & 0 \\ 0 & 0 & 0 & 0 \\ 1 & 0 & 1 & 0 \\ 0 & 0 & 0 & 0 \end{bmatrix}$$

$$M_{AP90} = \frac{1}{2} \begin{bmatrix} 1 & -1 & 0 & 0 \\ -1 & 1 & 0 & 0 \\ 0 & 0 & 0 & 0 \\ 0 & 0 & 0 & 0 \end{bmatrix} \qquad M_{AP135} = \frac{1}{2} \begin{bmatrix} 1 & 0 & -1 & 0 \\ 0 & 0 & 0 & 0 \\ -1 & 0 & 1 & 0 \\ 0 & 0 & 0 & 0 \end{bmatrix} \quad (14)$$

To determine the intensity measured by the PID for a given nanowire orientation $\theta$, we use

$$I_\theta = M_{AP}(\theta)S_{in} \quad (15)$$

where $I_\theta = [I_1\ I_2\ I_3\ I_4]^T$, and $S_{in} = [S_0\ S_1\ S_2\ S_3]^T$. We can represent the intensity measured at each $\theta$ as

$$I_0 = \frac{1}{2}(S_0 + S_1); \qquad I_{45} = \frac{1}{2}(S_0 + S_2);$$

$$I_{90} = \frac{1}{2}(S_0 - S_1); \qquad I_{135} = \frac{1}{2}(S_0 - S_2); \quad (16)$$

Using the four equations above, the Stokes parameters $S_0$, $S_1$, and $S_2$ can be expressed in terms of $I_\theta$:

$$S_0 = I_{0°} + I_{90°} \quad (17\text{-}1)$$

$$S_1 = I_{0°} - I_{90°} \quad (17\text{-}2)$$

$$S_2 = I_{45°} - I_{135°} \quad (17\text{-}3)$$

Note that the subscripts used here indicate the respective orientation of the superconducting nanowires.

**Appendix B:**

**The SDE$_{//}$ and SDE$_{\perp}$ vs. I$_b$ for the pixel with the highest PER**

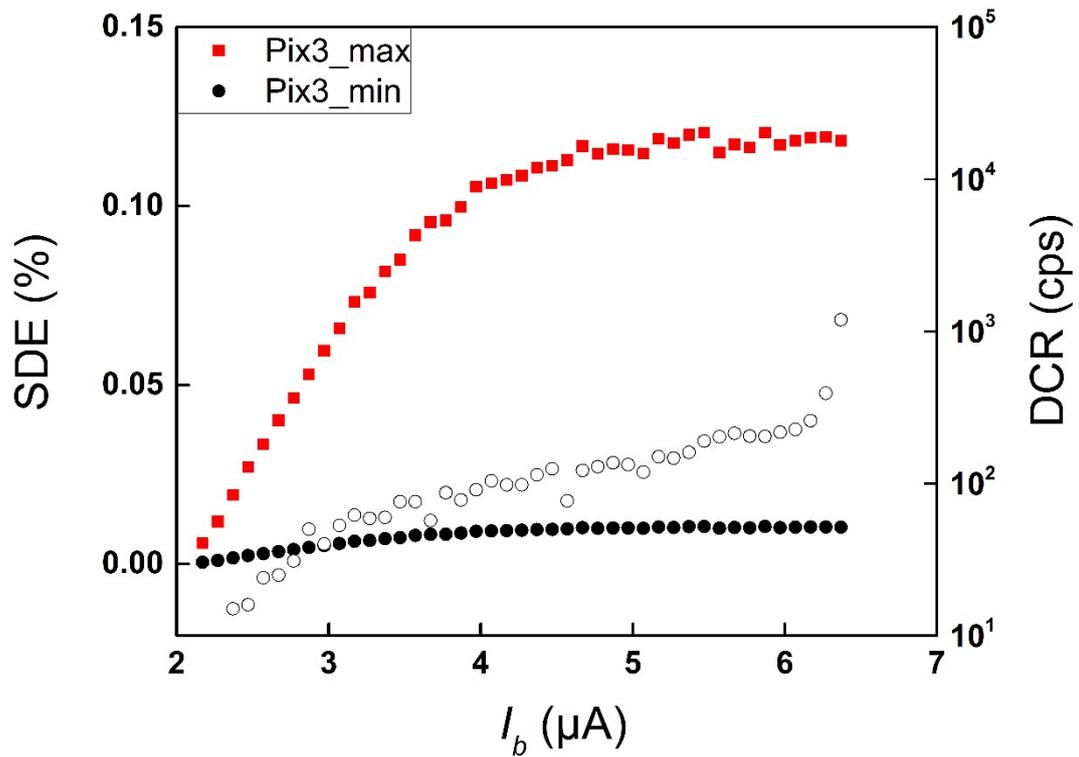

Figure 1. The SDE$_{//}$ and SDE of the single pixel with the highest PER as a function of bias current ($I_b$), and the DCR of the same pixel as a function of bias current ($I_b$).

**Appendix C:**

**The distribution of the PER for two devices**

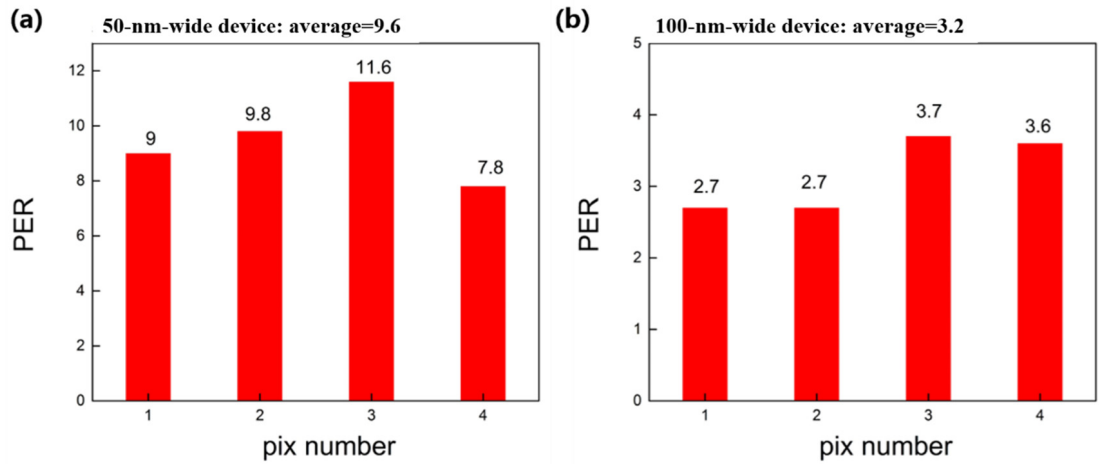

Figure 2. The distribution of the PER of the different width devices. The average PER of the 50nm wide SNSPD is 9.6, whereas that of the 100nm wide one is 3.2. The differences in PER values between the nanowire pixels could be attributed to variations in fabricated wire width, since the PER is sensitive to changes in wire width. For example, according to the simulation shown in Fig. 1(e) of the main text, if the width of the nanowire varies between 45nm to 50nm, the corresponding PER would change from 21 to 17, i.e., a reduction of ~4 in the PER value. This relative reduction is consistent with the experimental data shown in Fig. 2 (a), where the largest PER reduction is approximately 3.8. Similar conclusions can also be drawn in the 100-nm-wide nanowire case. The simulated PER value is higher than the experimental value, possibly attributed to imperfect fabrication of the nanowire geometrical structure [43] and use of inaccurate optical parameters in the simulation.

Applied Physics Letters **101**(2012).